\documentclass[twocolumn,prb,amsmath,amssymb,showpacs,superscriptaddress]{revtex4-1}

\usepackage{graphicx}
\usepackage[hypertex]{hyperref}

\renewcommand{\Re}{\mathop\mathrm{Re}\nolimits}
\renewcommand{\Im}{\mathop\mathrm{Im}\nolimits}

 \DeclareMathOperator{\tr}{tr}

\begin{document}

\title{Surface impedance of superconductors with weak magnetic impurities}

 \author{Ya.~V.~Fominov}
 \affiliation{L.~D.~Landau Institute for Theoretical Physics RAS, 119334 Moscow, Russia}

 \author{M.~Houzet}
 \affiliation{SPSMS, UMR-E 9001, CEA-INAC/UJF-Grenoble 1, F-38054 Grenoble, France}

 \author{L.~I.~Glazman}
 \affiliation{Departments of Physics, Yale University, New Haven, CT 06520, USA}

\date{27 December 2011}

\begin{abstract}
Magnetic impurities affect the spectrum of excitations of a superconductor and thus influence its impedance. We concentrate on
the dissipative part of the surface impedance. We investigate its dependence on frequency, the density and strength of magnetic impurities, and the density and temperature of quasiparticles. Even a small concentration of weak magnetic impurities significantly modifies the excitation spectrum in the vicinity of the BCS gap. Therefore, we give special attention to the absorption threshold behavior at zero temperature and to the low-frequency absorption by quasiparticles. The discrete energy states introduced at low density of magnetic impurities may serve as traps for nonequilibrium quasiparticles, reducing the absorption in some range of low radiation frequencies.
\end{abstract}


%
%
%

\maketitle

\section{Introduction}

Electron scattering off magnetic impurities, unlike the potential scattering, substantially modifies the properties of \textit{s}-wave superconductors. It was realized in the seminal work by Abrikosov and Gor'kov\cite{AG} (AG) that the presence of magnetic impurities reduces the superconducting gap below its BCS value and eventually may lead to the phenomenon of gapless superconductivity. The gap suppression was investigated by means of tunneling between normal and superconducting electrodes.\cite{Exp-Woolf,Exp-Edelstein} Beyond the AG theory, it was realized that a single magnetic impurity creates a localized state within the BCS gap.\cite{Yu,Soda,Shiba,Rusinov} A small concentration of magnetic impurities results in the formation of an impurity band\cite{Shiba,Rusinov} which merges with the continuum as the concentration exceeds a certain value (see Ref.~\onlinecite{BVZ} for a review). (The original AG theory is valid above that value.) The impurity band was investigated by tunneling experiments with alloys (such as PbMn)\cite{Exp-Bauriedl} and normal metal--superconductor bilayers,\cite{Exp-Dumoulin} and also by means of thermal transport in superconducting films.\cite{Exp-Ginsberg} The observation of discrete levels associated with a separate magnetic impurity came later in the scanning tunneling microscopy experiments.\cite{Exp-Yazdani,Exp-Xue}

The effect of magnetic impurities came under scrutiny recently in connection with limits on performance of superconducting elements in qubits\cite{Clarke} and photon detectors.\cite{Klapwijk} In these applications, superconductors should perform in the ac regime. There is also some experimental effort directly aimed at observation of the effect of magnetic impurities on the surface impedance of superconducting multilayers\cite{Palomba} and unconventional superconductors.\cite{Srikanth,Prokhorov,Hein} Theoretically, the results on impedance\cite{Palomba,Srikanth,Prokhorov,Hein} were analyzed in the framework of a simple two-fluid model. The two-fluid model assumes a simplified frequency-independent Drude-type dissipative conductivity $\sigma$, while the influence of magnetic impurities on the electron spectrum should lead to a nontrivial $\sigma(\omega)$ dependence on the frequency $\omega$.

A theoretical approach to studying the surface impedance in superconductors was developed in the celebrated work of Mattis and Bardeen,\cite{MB} which treated the anomalous skin effect (the skin depth is smaller than the mean free path) taking nonmagnetic disorder into account. Independently, the clean case was considered by Abrikosov, Gor'kov, and Khalatnikov\cite{AGKh,AGD} in the Green-function language. Later, Nam\cite{Nam} employed the same technique and demonstrated that the extreme anomalous limit is equivalent (up to the expressions for the normal-metallic conductivities) to the local limit, realized in the dirty regime. The complex conductivity of a superconductor containing both potential and magnetic impurities was addressed by Skalski \textit{et al.}\cite{Skalski} Their paper does contain general results applicable in the case of weak magnetic scattering; however, similarly to the AG theory, scattering off magnetic impurities was considered in the Born approximation, which did not allow to treat the effect of localized states on the impedance. In Ref.~\onlinecite{LO_1971}, Larkin and Ovchinnikov considered superconductors with inhomogeneous pairing, which under certain assumptions is formally equivalent to the presence of magnetic impurities in the AG regime. Some of their results can therefore be applied to this problem.
However, a convenient formulation of the theory, suitable for the experimentally relevant dirty limit and making it possible to treat both the AG regime and the impurity band, as well as detailed quantitative investigation of the effect of magnetic impurities on the surface impedance in different regimes over temperature, frequency, and electron scattering rate off magnetic impurities is still lacking, to the best of our knowledge.

In this paper, we study the influence of magnetic impurities on the complex conductivity and surface impedance of a superconductor. We are especially interested in the limit of low frequency of the impinging radiation and low temperature, and in the effect of the impurity band on the dissipation. Quantitative results are presented for the experimentally relevant dirty limit (with respect to the nonmagnetic disorder).

The paper is organized as follows. First, in Sec.~\ref{sec:clean}, we consider the limit of weak scattering off magnetic impurities and disregard nonmagnetic disorder (``clean'' limit). In that limit, we find the localized states bound to magnetic impurities and discuss the formation of the impurity band at finite but small concentration of magnetic impurities and the formation of the AG density of states at higher concentrations. We also present there a qualitative consideration of the dissipative conductivity. The main features of the developed simple picture carry over to the realistic case of a ``dirty'' superconductor doped with magnetic impurities. In
Sec.~\ref{sec:dirty}, we consider the dirty limit for the nonmagnetic disorder in the framework of the quasiclassical formalism of the Usadel equation. We first obtain the spectrum of quasiparticle excitations in the presence of magnetic impurities (treated beyond the Born approximation).
Then we address the effect of magnetic impurities on the low-temperature dissipative conductivity. In addition to the low-frequency limit, we also consider in detail the frequency dependence of the dissipation near the thresholds determined by the gaps in the excitation spectrum. Our results are summarized in Sec.~\ref{sec:concl}. Some technical details are presented in the Appendixes.

\section{Clean limit} \label{sec:clean}

To highlight the main features of the problem related to the impurity states, we consider here a simplified case of a superconductor with magnetic impurities only.
In addition, considering the hybridization of single-impurity states into an impurity band we first assume magnetic impurities to be polarized (magnetic moments pointing up the $z$ axis), while the impurity positions are random.
Furthermore, we assume the exchange constant $J$ of interaction of localized classical spins $S$ with the spins of itinerant electrons to be small. Introducing the dimensionless parameter
\begin{equation}
\zeta=\pi\nu_0 JS,
\end{equation}
describing the ``strength'' of a single magnetic impurity, we can write the latter requirement as
\begin{equation}
\zeta\ll 1.
\end{equation}
Here $\nu_0 = k_F^2/2 \pi^2 v_F$ is the electron density of states (DoS)
per spin projection at the Fermi level in the normal state, with $k_F$ and $v_F$ being the Fermi wave vector and the Fermi velocity, respectively. Weak exchange interaction will modify substantially the excitation spectrum only close to the gap. Concentrating on that energy region, we simplify the Hamiltonian of the system in the following way. First, we express the operator of exchange interaction of electrons with an impurity (placed at origin, $\mathbf R=0$) in terms of the Bogoliubov quasiparticles ($\psi$ and $\alpha$ are the operators of electrons and Bogoliubov quasiparticles, respectively):
\begin{multline}
J S \sum_{\mathbf k_1,\mathbf k_2,\sigma} \sigma \psi_{\mathbf k_1 \sigma}^\dagger \psi_{\mathbf k_2 \sigma} \\
= J S \sum_{\mathbf k_1,\mathbf k_2} \left[ (u_{\mathbf k_1} u_{\mathbf k_2} +v_{\mathbf k_1} v_{\mathbf k_2}) (\alpha_{\mathbf k_1 \uparrow}^\dagger \alpha_{\mathbf k_2 \uparrow} -\alpha_{\mathbf k_1 \downarrow}^\dagger \alpha_{\mathbf k_2 \downarrow}) \right. \\
\left. - (u_{\mathbf k_1} v_{\mathbf k_2} -u_{\mathbf k_2} v_{\mathbf k_1}) (\alpha_{\mathbf k_1 \uparrow}^\dagger \alpha_{-\mathbf k_2 \downarrow}^\dagger -\alpha_{-\mathbf k_1 \downarrow} \alpha_{\mathbf k_2 \uparrow}) \right],
\end{multline}
where $\mathbf k_{1(2)}$ are momenta and $\sigma$ is the spin index
($+$ for $\uparrow$, $-$ for $\downarrow$). At the same time, the standard BCS part of the Hamiltonian becomes diagonal after the Bogoliubov transformation, with the excitation energies $E_{\mathbf k} =\sqrt{\xi_{\mathbf k}^2 +\Delta^2}$ in the superconducting state. Here $\Delta$ is the BCS gap and $\xi_{\mathbf k}$ are the excitation energies in the normal state.
Close to the gap ($|E_{\mathbf k}-\Delta|\ll\Delta$), we can approximate $E_{\mathbf k} \approx \Delta + \xi_{\mathbf k}^2/2\Delta$, while the Bogoliubov coherence factors are simply $u_{\mathbf k} \approx v_{\mathbf k} \approx 1/\sqrt 2$, so finally the total Hamiltonian takes the form
\begin{equation} \label{Ham}
\hat H = \sum_{\mathbf k, \sigma} \varepsilon_{\mathbf k} \alpha_{\mathbf k \sigma}^\dagger \alpha_{\mathbf k \sigma} + J S \sum_{\mathbf k_1,\mathbf k_2,\sigma} \sigma \alpha_{\mathbf k_1 \sigma}^\dagger \alpha_{\mathbf k_2 \sigma},
\end{equation}
where $\varepsilon_{\mathbf k} = \xi_{\mathbf k}^2/2\Delta$ and we measure the quasiparticle energies from the edge of the continuous spectrum $\Delta$:
\begin{equation}
\varepsilon= E -\Delta.
\end{equation}

\subsection{Single-impurity bound state}

It is easy to see that Hamiltonian (\ref{Ham}) has a bound state for quasiparticles with spin $\sigma$ satisfying the condition $JS\sigma<0$. For definiteness, hereafter we set $J>0$; then the bound state exists for spin down. We denote the wave function of the bound state in the momentum representation as $\varphi_{\mathbf k}$; the Schr\"odinger equation for it, which follows from the form of Hamiltonian (\ref{Ham}), reads
\begin{equation} \label{eqphi}
(\varepsilon_0-\varepsilon_{\mathbf k})\varphi_{\mathbf k}= -JS\sum_{\mathbf k_1} \varphi_{\mathbf k_1}.
\end{equation}
Clearly, the wave function must have the form $\varphi_{\mathbf k}\propto (\varepsilon_0-\varepsilon_{\mathbf k})^{-1}$. Substituting it into Eq.\ (\ref{eqphi}), we find the equation for the energy of the bound state ($\varepsilon_0<0$),
\begin{equation} \label{Sch_E}
1=JS\int\frac{d\mathbf k}{(2\pi)^3}\frac 1{\varepsilon_{\mathbf k}-\varepsilon_0}\,,
\end{equation}
yielding
\begin{equation} \label{bs}
\varepsilon_0=-2\zeta^2 \Delta
\end{equation}
at $\zeta\ll 1$. The power-law dependence, $\varepsilon_0\propto\zeta^2$, is a direct consequence of the excitations spectrum $\varepsilon_{\mathbf k}$. The convergence of the integral in Eq.\ (\ref{Sch_E}) at $\xi_{\mathbf k}\sim\zeta\Delta\ll\Delta$ validates the use of the simplified form of $E_{\mathbf k}$. We emphasize that in order to obtain the bound state, it was essential to go beyond the Born approximation.\cite{Shiba,Rusinov} The wave function of the bound state in the coordinate representation has the form
\begin{equation} \label{wf}
\varphi (\mathbf r) = \frac{\sin k_F r}{r\sqrt{\pi \xi_0/2\zeta}}\exp\left(-\frac r{\xi_0/2\zeta}\right),
\end{equation}
where $\xi_0 = v_F/\Delta$ is the coherence length of a clean superconductor.

The wave function of the localized state decays exponentially on the spatial scale which exceeds the coherence length by the factor $1/\zeta$. Clearly, the effect of multiple magnetic impurities on the quasiparticle spectrum depends crucially on the characteristic distance between them compared to $\xi_0/\zeta$. At lowest concentrations $n_s$, separate magnetic impurities create a $\delta$-peak in the quasiparticle DoS. With the increase of the concentration the peak broadens. As long as the average inter-impurity distance is large, the broadening is due to rare occurrences of closely located impurity pairs producing split levels due to their hybridization.\cite{LGP} However, at higher concentrations, $n_s (\xi_0/\zeta)^3\gg 1$, the bound state on a given impurity overlaps with a large number of other localized states and forms a well-defined impurity band. This large parameter allows one to treat the DoS in that band self-consistently.

\subsection{Impurity band} \label{subsec:imp-band}

We start with an estimate of the impurity band width $\mathcal W$. A quasiparticle initially localized on a given impurity may hop on $\sim n_s(\xi_0/\zeta)^3\gg 1$ ``nearest neighbors''. Because of their large number, the return of the quasiparticle is improbable. Therefore, we may introduce the escape rate $\Gamma_i$ for a given impurity. It will determine the level width, and we will associate it with the width of the band, $\Gamma_i\sim \mathcal W$. On the other hand, the escape rate from a given site is proportional to the DoS on the ``receiving'' ones. The latter is inversely proportional to $\mathcal W$. This way, the formula for the escape rate turns into a self-consistent equation for $\mathcal W$ (this is the essence of the self-consistent Born approximation for degenerate states, analyzed in detail in the context of the broadening of high Landau levels\cite{Ando,Chalker}). To implement the scheme, we denote the hopping matrix element between two impurity states $i$ and $j$, at positions $\mathbf R_{i}$ and $\mathbf R_{j}$,
 as $t_{ij}$. The said self-consistency equation reads
\begin{equation} \label{W}
\mathcal W \sim \sum_j |t_{ij}|^2\cdot\frac 1{\mathcal W}
\end{equation}
(upon summing over positions $j$, the dependence on $i$ disappears). The estimate of $t_{ij}$ should be obtained from a consideration of the two-impurity problem. Tunneling between two impurity sites splits the degenerate level in two. The resulting energy splitting depends on the inter-impurity distance $R_{ij}$. Similarly to the conventional tight-binding problem, it is this energy splitting that should be identified with $2|t_{ij}|$. This way, we find  $|t_{ij}|\propto |\varphi (\mathbf R_{ij})|$. To estimate the proportionality coefficient here, we note that the energy splitting reaches a value of the order of $|\varepsilon_0|$ if $k_F R_{ij} \sim 1$. Therefore,
$|t_{ij}|\sim |\varepsilon_0|\cdot \left| \varphi (\mathbf R_{ij}) \right|/ \left| \varphi (1/k_F) \right|$.
This estimate is confirmed by a solution of the energy splitting problem, which is easily obtained by the generalization of Eqs.\ (\ref{Ham})--(\ref{Sch_E}) to the case of two impurities.\cite{twoimp}

Summation over the random positions of the impurities $j$ results in $\sum_j |t_{ij}|^2 \sim \varepsilon_0^2 n_s /|\varphi (1/k_F)|^2\sim\varepsilon_0^2 n_s \xi_0/\zeta k_F^2$. Remarkably, we used here only the value of the wave function $\varphi (\mathbf r)$ close to the impurity and the fact that $\varphi (\mathbf r)$ is normalized. This is why one may expect that the obtained estimate of $\sum_j |t_{ij}|^2$ remains valid in the presence of nonmagnetic disorder, as long as the corresponding mean free path is large compared to the electron Fermi wave length.
(This expectation is indeed confirmed by the rigorous consideration
of Sec.~\ref{sec:dirty}.) Substituting the estimate of $\sum_j |t_{ij}|^2$ into Eq.\ (\ref{W}) and replacing $k_F^2/v_F\sim\nu_0$, we find expression for $\mathcal W$ up to a numerical factor. Together with this factor (which will be found later),
the bandwidth in terms of measurable quantities $n_s$ and $\varepsilon_0$, reads
\begin{equation} \label{Wfin}
\mathcal W = 4\frac{2^{1/4}}{\pi^{1/2}} \left(\frac{n_s}{\nu_0\Delta}\right)^{1/2} \left( \frac\Delta{|\varepsilon_0|} \right)^{1/4} |\varepsilon_0|.
\end{equation}
This expression is valid when the band is narrow, $\mathcal W\ll |\varepsilon_0|$.

\begin{figure}[t]
 \includegraphics[width=\hsize]{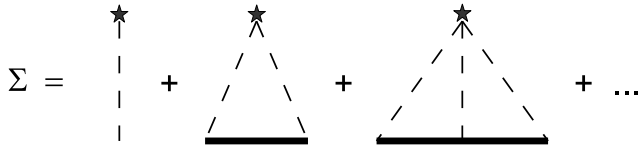}
\caption{Self-energy $\Sigma(\mathbf k,\varepsilon)$ corresponding to the self-consistent $T$-matrix approximation. The scattering on a single impurity (dashed lines) is treated in all orders of the perturbation series. The self-consistency is achieved by inserting the bold lines for the Green functions, which means that they are calculated with the help of the presented self-energy.}
 \label{fig:SCBA}
\end{figure}

The next question is the DoS $\nu_B(\varepsilon)$ for the Bogoliubov quasiparticles (hence the $B$ subscript) inside the impurity band which can be found from the momentum-integrated Green function $g(\varepsilon)$ as
\begin{equation}
\nu_B(\varepsilon) = -\frac 1\pi \Im g(\varepsilon+i0),\qquad g(\varepsilon) = \int \frac{d\mathbf k}{(2\pi)^3} G_B(\mathbf k,\varepsilon).
\end{equation}
The Green function $G_B(\mathbf k,\varepsilon)$ of the Bogoliubov quasiparticles is determined by the Hamiltonian, which now, in contrast to Hamiltonian (\ref{Ham}), contains magnetic impurities of finite concentration:
\begin{equation} \label{Ham1}
\hat H = \sum_{\mathbf k, \sigma} \varepsilon_{\mathbf k} \alpha_{\mathbf k \sigma}^\dagger \alpha_{\mathbf k \sigma} + J S \sum_{\mathbf k_1,\mathbf k_2,\sigma,j} \sigma \alpha_{\mathbf k_1 \sigma}^\dagger \alpha_{\mathbf k_2 \sigma} e^{i(\mathbf k_2-\mathbf k_1) \mathbf R_j} .
\end{equation}
The Green function can be written in terms of the self-energy $\Sigma(\mathbf k,\varepsilon)$ and we treat the self-energy within the self-consistent $T$-matrix approximation,\cite{AS} which diagrammatically amounts to the geometric series of Fig.~\ref{fig:SCBA}:
\begin{gather}
G_B(\mathbf k,\varepsilon) = \frac 1{\varepsilon -\varepsilon_{\mathbf k} - \Sigma(\mathbf k,\varepsilon)}, \label{G} \\
\Sigma(\mathbf k,\varepsilon) = -\frac{n_s J S}{1+J S \int \frac{d\mathbf q}{(2\pi)^3} G_B(\mathbf q,\varepsilon)}
\end{gather}
(the sign in front of $J$ in the self-energy corresponds to quasiparticle spin down, which is of interest for us).
Integrating Eq.\ (\ref{G}) over momentum, we obtain the self-consistency condition for $g_\downarrow(\varepsilon)$:
\begin{equation} \label{g}
JS g_\downarrow = - \sqrt{\frac{|\varepsilon_0|}{-\varepsilon -\frac{n_s JS}{1+JSg_\downarrow}}}.
\end{equation}
The impurity band corresponds to nonzero DoS, that is, to the nonzero imaginary part of $g_\downarrow$, in some domain of negative $\varepsilon$. Analyzing Eq.\ (\ref{g}) in the limit $n_s \ll \zeta \nu_0 \Delta$ (the limit of narrow band), we find a semicircle DoS inside the impurity band:
\begin{equation} \label{nuom}
\nu_{B\downarrow}(\varepsilon) = \frac{4 n_s}{\pi \mathcal W} \Re \sqrt{1 - \left( \frac{\varepsilon+|\varepsilon_0|}{\mathcal W/2} \right)^2},
\end{equation}
which is centered at $\varepsilon_0$, found earlier in Eq.\ (\ref{bs}), and has the width given by Eq.\ (\ref{Wfin}) [note that it is the calculation leading to Eq.\ (\ref{nuom}) that establishes the numerical factor in Eq.\ (\ref{Wfin})].
The total DoS inside the impurity band is equal to $n_s$, demonstrating that each impurity brings one state into the system.\cite{Zittartz}
The same Eq.\ (\ref{g}) can be employed to demonstrate that the lower edge of the continuous spectrum
shifts from zero to $\varepsilon_c = n_s^2/8\pi^2\nu_0^2\Delta$ at finite impurity concentrations, and the behavior of the DoS near the edge is
\begin{equation}
\nu_{B\downarrow}(\varepsilon) = \nu_0 \frac{\sqrt{2\Delta (\varepsilon-\varepsilon_c)}}{\varepsilon_c}.
\end{equation}
(The positive value of $\varepsilon_c$ is a manifestation of the level repulsion between the states of the continuum and the impurity states.)

As we mentioned before, a well-defined impurity band exists [and can be described by the self-consistent $T$-matrix method leading to DoS (\ref{nuom})] if $n_s (\xi_0/\zeta)^3\gg 1$, which translates into $\mathcal W\gg |\varepsilon_0|\sqrt{\zeta}(\Delta/E_F)$, where $E_F$ is the Fermi energy. With the increase of the impurity concentration, the impurity band broadens, $\mathcal W\propto\sqrt n_s$. At $n_s \sim \zeta\nu_0\Delta$, the width of the band becomes of the same order as the distance from the center of the band to the continuum edge, resulting in the merger of the impurity band with the continuum of quasiparticle states. [Quantitatively, the analysis of Eq.\ (\ref{g}) shows that the energy gap between the impurity band and continuum
exists if $n_s/(\zeta\nu_0\Delta)<16\pi/27\approx 1.86$.] The two conditions for the existence of the impurity band, $n_s (\xi_0/\zeta)^3\gg 1$ and $n_s \lesssim \zeta\nu_0\Delta$, are consistent since $\nu_0 \Delta \xi_0^3/\zeta^2 \sim (E_F/\Delta \zeta)^2 \gg 1$.

\begin{figure}[t]
 \includegraphics[width=\hsize]{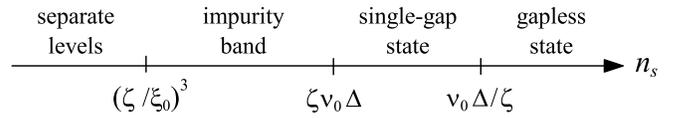}
\caption{Different regimes for structure of the DoS due to polarized magnetic impurities of concentration $n_s$.}
 \label{fig:Axis_ns}
\end{figure}

In the above discussion, we did not treat the self-consistency equation explicitly, assuming $\Delta$ that we used to be the self-consistent order parameter corresponding to the given impurity parameters.
Actually, $\Delta$ is suppressed by magnetic impurities.
A gapless state is realized at a critical concentration
$n_s\sim \nu_0\Delta/\zeta$. At that concentration the average Zeeman energy associated with the magnetic impurities,
\begin{equation}
E_z = n_s JS,
\end{equation}
is comparable to $\Delta$.\cite{Clogston,Chandrasekhar}
Figure~\ref{fig:Axis_ns} illustrates the results of the above discussion.

The above consideration of spin-polarized magnetic impurities can be easily generalized to the case of randomly oriented spins (see Appendix~\ref{app:relation}). It yields Eq.\ (\ref{app:nuom3}) for the DoS --- now summed over the quasiparticle spin directions --- inside an impurity band of width $W$ [see Eq.\ (\ref{app:Wfin3})]. Comparing Eqs.\ (\ref{app:nuom3}) and (\ref{app:Wfin3}) with Eqs.\ (\ref{nuom}) and (\ref{Wfin}), we  notice the relation $W = \mathcal W /\sqrt 2$ between the widths of the band. Indeed, randomness of impurity spin orientations leads to an additional $\cos(\vartheta/2)$ factor in the hopping matrix element $t_{ij}$ between two impurity states [see Eq.\ (\ref{W})], and hence to the $1/2$ factor in the average $\left< | t_{ij} |^2 \right>$, finally changing the width of the band by the $1/\sqrt 2$ factor. (Here $\vartheta$ is the angle between the spins of the two impurities.)

Another difference of the random-spin case from the spin-polarized case is the absence of the average Zeeman energy. Upon increasing $n_s$, the gapless state is then reached due to the electron spin-flip scattering off magnetic impurities.\cite{AG} The latter is characterized by a scattering rate $1/\tau_s$ which, in the Born approximation, is
\begin{equation}
\frac 1{\tau_s} = 2\pi\nu_0 J^2 S^2 n_s= \frac{2\zeta^2 n_s}{\pi\nu_0}.
\end{equation}
The transition occurs when $1/{\tau_s} = \Delta$, that is, at
a concentration of magnetic impurities $n_s\sim \nu_0\Delta/\zeta^2$.

\subsection{Effect of the impurity band on the electromagnetic field absorption}

At finite but low temperatures $T$, quasiparticles first populate the impurity band. This results in a finite dissipative part $\sigma_1(\omega)$ of conductivity at arbitrary low frequencies. We estimate $\sigma_1(\omega)$ at $\omega\ll W$ using a suitable version\cite{deGennes} of the Mattis-Bardeen theory.\cite{MB} We disregard possible spin selection rules\cite{Littlewood,SBS}
by assuming that the spins of magnetic impurities are randomly oriented.

We may relate $\sigma_1$ to the absorption power per unit volume, $w=(1/2)\sigma_1 \mathcal E^2$, where $\mathcal E$ is the amplitude of the applied ac electric field. The absorption power can be evaluated as the product of the photon energy quantum $\omega$ and the rate of creation of quasiparticles. The latter is $\propto \mathcal E^2$ and is evaluated with the help of the Fermi golden rule. The perturbation leading to the excitation of the system is $B=(e/2)(\mathbf A \hat{\mathbf v}+\hat{\mathbf v} \mathbf A)$, where $\hat{\mathbf v}$ is the operator of velocity of an electron and $\mathbf A$ is the vector potential of the ac field; its amplitude is $A=\mathcal E/\omega$. The absorption power associated with the photon-induced transitions of quasiparticles within the impurity band equals\cite{deGennes}
\begin{multline} \label{abspow2}
w=2\pi\omega \overline{B^2}\int dE_1 dE_2 [n(E_2)-n(E_1)] \\
\times \nu_B (E_1) \nu_B (E_2) (u_1 u_2+v_1 v_2)^2 \delta (E_1-E_2-\omega).
\end{multline}
Writing Eq.\ (\ref{abspow2}), we accounted for the fact that the states in the impurity band do not correspond to a defined momentum, and consequently the latter is not conserved in the transitions. To estimate the average over the states $\overline{B^2}$, we employ the arguments of the work of Mattis and Bardeen,\cite{MB} which allow one to express it in terms of an analog of the normal conductivity for the impurity band. That leads us to the estimate $\overline{B^2}\sim {\mathcal E}^2(ev_F/\omega)^2(\tau/\nu)$, where $\tau$ is the relaxation time and $\nu$ is a characteristic value of the DoS in the impurity band. Substituting $\tau\sim 1/W$ and $\nu\sim n_s/W$ [see Eq.\ (\ref{app:nuom3})], we find
\begin{equation} \label{B2}
\overline{B^2}\sim \frac 1{n_s} \left( \frac{e v_F}\omega \right)^2 {\mathcal E}^2.
\end{equation}
In the case of ``shallow'' levels ($\zeta\ll 1$) we may also replace the combination of the coherence factors,
$(u_1 u_2+v_1 v_2)^2 \sim 1$. At lowest frequencies, $\omega\ll T,W$, we may expand the difference of the quasiparticle distribution functions in Eq.\ (\ref{abspow2}) as $n(E_2)-n(E_1)\approx -(dn/dE)\omega$. As a result, we find frequency-independent asymptotic forms for the dissipative conductivity at $T\ll W$ and $T\gg W$, which we express in terms of the density $n_\mathrm{qp}^{(b)}$ of quasiparticles in the impurity band [hence the $(b)$ superscript]:
\begin{equation}
\label{eq:sigmab-clean1}
 \sigma_1^{(b)} \sim
 \left\{
\begin{array}{ll}
n_\mathrm{qp}^{(b)} (e v_F)^2/(T W^3)^{1/2}, & \text{at } \omega \ll T\ll W, \\
n_\mathrm{qp}^{(b)} (e v_F)^2/(T W) , & \text{at } \omega\ll W\ll T.
\end{array} \right.
\end{equation}
In equilibrium, the  density of quasiparticles is
\begin{align}
n_\mathrm{qp}^{(b)} &= 2\int dE \nu_B(E) n(E) \\
&\sim \left\{
\begin{array}{ll}
 n_s \left( T/W \right)^{3/2} e^{-(\Delta-|\varepsilon_0|- W/2)/T}, & \text{at } T\ll W,\\
 n_s e^{-(\Delta-|\varepsilon_0|)/T}, & \text{at } T\gg W.
\end{array} \right.
\nonumber
\end{align}
In the opposite limit of photon energies high compared to temperature, we may neglect $n(E_1)$ in Eq.\ (\ref{abspow2}), finally finding
\begin{equation}
\label{eq:sigmab-clean2}
\sigma_1^{(b)}(\omega) \sim n_\mathrm{qp}^{(b)} (e v_F)^2/(\omega W^3)^{1/2} ,\quad T\ll\omega\ll W.
\end{equation}

In an ideal superconductor at zero temperature, photon absorption is possible only for photons with energy $\omega$ exceeding the threshold $2\Delta$ (each photon creates a pair of quasiparticles). The presence of states within the gap $\Delta$ reduces the threshold frequency and modifies the dependence of the dissipative part of conductivity $\sigma_1$ on frequency for $\omega$ exceeding the threshold.
Similarly to the above consideration, one may relate $\sigma_1(\omega)$ to the absorption power (cf.\ Ref.~\onlinecite{deGennes}),
\begin{multline} \label{abspow3}
w=2\pi\omega \overline{B^2} \int dE_1 dE_2\\
\times \nu_B (E_1) \nu_B (E_2) (u_1 v_2- v_1 u_2)^2 \delta (E_1+E_2-\omega).
\end{multline}
As follows from Eq.\ (\ref{app:nuom3}), absorption is possible only if the frequency exceeds the threshold,
\begin{equation} \label{omth}
\omega>\omega_\mathrm{th}\,,\quad \omega_\mathrm{th} =2\left( \Delta-|\varepsilon_0|- \frac{W}2 \right).
\end{equation}
To evaluate the absorption, however, the accepted approximation of shallow levels is insufficient, because it yields $(u_1 v_2- v_1 u_2)^2=0$. As follows from the detailed calculation presented in Sec.~\ref{sec:dissip_zeroT} this factor $(u_1 v_2- v_1 u_2)^2\sim\zeta^2$; the dissipative conductivity scales as $\sigma_1^{(b)}(\omega)\propto (\omega-\omega_\mathrm{th})^2$ near the threshold.

\section{Dirty limit} \label{sec:dirty}

We turn now to the practically important case of short elastic mean free time $\tau_p$ caused by potential scattering of electrons and consider the dirty limit, $\Delta\ll 1/\tau_p$.
We will see that potential disorder does not affect the electron spectrum.\cite{MS} However, its role is essential for transport properties (complex conductivity and impedance).

In the dirty limit, magnetic impurities of small concentration can be described in the framework of the Usadel equation.\cite{Usadel} The form of the Usadel equation, widely used in literature,\cite{LO,RS} corresponds to the Born limit for magnetic scattering  (with random positions and orientations of magnetic impurities). At the same time, the generalization to arbitrary strength of magnetic scattering can also be obtained, as shown by Marchetti and Simons.\cite{MS} Employing the standard $\theta$-parametrization of the normal and anomalous quasiclassical Green functions,
\begin{equation} \label{GFtheta}
G=\cos\theta,\qquad F=\sin\theta,
\end{equation}
we can write the generalized form as\cite{MS,footnote_sc}
\begin{equation} \label{Usadel_mod}
\frac D2 \nabla^2 \theta + iE \sin\theta + \Delta \cos\theta -\frac 1{2\tau_s} \frac{\sin 2\theta}{1+ \zeta^4 +2\zeta^2 \cos 2\theta} =0,
\end{equation}
where $D=v_F^2\tau_p/3$ is the diffusion constant and $E$ is the electron energy measured from the Fermi level.
We are interested in the homogeneous solution. Following Ref.~\onlinecite{MS}, we define a ``renormalized'' energy and order parameter,
\begin{align}
\tilde E &= E +\frac i{2\tau_s} \frac{\cos\theta}{1+ \zeta^4 +2\zeta^2 \cos 2\theta}, \label{tildeE} \\
\tilde\Delta &= \Delta -\frac 1{2\tau_s} \frac{\sin\theta}{1+ \zeta^4 +2\zeta^2 \cos 2\theta}. \label{tildeD}
\end{align}
With these quantities, the homogeneous Usadel equation becomes simple,
\begin{equation}
i\tilde E \sin\theta + \tilde\Delta \cos\theta =0,
\end{equation}
and can be formally solved as
\begin{equation} \label{tEtD}
\cos\theta = \frac u{\sqrt{u^2-1}},\quad \sin\theta = \frac i{\sqrt{u^2-1}},
\end{equation}
where $u=\tilde E/\tilde\Delta$. At the same time, this is not a solution yet, but an alternative parametrization of the Green functions in terms of $u$. The Usadel equation actually determines the Green functions' dependence on energy $E$; hence, in Eq.\ (\ref{tEtD}) $u$ itself depends on $E$. The latter dependence is determined by the equation, directly following from Eqs.\ (\ref{tildeE}) and (\ref{tildeD}):
\begin{equation} \label{u_eq}
u \left( 1+\gamma_s \frac{\sqrt{1-u^2}}{u^2-\epsilon_0^2} \right) = \frac E\Delta,
\end{equation}
where
\begin{equation} \label{gammaepsilon}
\gamma_s = \frac 1{(1+\zeta^2)^2} \frac 1{\tau_s \Delta},\qquad \epsilon_0 = \left| \frac{1-\zeta^2}{1+\zeta^2} \right|.
\end{equation}
Here $\gamma_s$ is the exact scattering rate due to magnetic impurities, normalized by $\Delta$, and $\epsilon_0$ can be shown to determine the center of the impurity band, also in units of $\Delta$. Since we focus on the case when each impurity is weak, $\zeta\ll 1$, we simplify the above expressions as
\begin{equation} \label{eps0}
\gamma_s = \frac 1{\tau_s \Delta},\qquad \epsilon_0 = 1-2\zeta^2.
\end{equation}

Once found, $u$ can be immediately applied for analyzing the spectrum. The DoS per spin is given by
\begin{equation} \label{DoS_u}
\frac{\nu(E)}{\nu_0} = \Re G(E) = \Re \frac u{\sqrt{u^2-1}}.
\end{equation}
It turns out that the potential scattering does not influence the energy spectrum, since Eqs.\ (\ref{u_eq}), (\ref{gammaepsilon}), and (\ref{DoS_u}) coincide\cite{MS} with those obtained by Shiba\cite{Shiba} and Rusinov\cite{Rusinov} in the clean limit (see also Appendix~\ref{app:relation}).

\subsection{Green functions and the density of states} \label{sec:DoS}

Finite DoS corresponds to complex solutions of Eq.\ (\ref{u_eq}) for $u$. Analyzing the behavior of the function in the left-hand side (l.h.s.) of Eq.\ (\ref{u_eq}) at $0<u<1$, Shiba\cite{Shiba} found the domain of energies [the right-hand side (r.h.s.) of Eq.\ (\ref{u_eq})] that cannot be matched by \textit{real} values of $u$. At positive energy, this domain is the impurity band centered around $E_0=|\varepsilon_0|\Delta$. Calculating $u$ inside the narrow band, we find
\begin{equation} \label{nuom3}
\nu(E) = \frac{n_s}{\pi W} \Re \sqrt{1 - \left( \frac{E-E_0}{W/2} \right)^2},\quad \text{at } E>0,
\end{equation}
for the DoS and
\begin{equation} \label{Wfin3}
W = 4\frac{1}{2^{1/4}\pi^{1/2}} \left(\frac{n_s}{\nu_0\Delta}\right)^{1/2} \left( \frac\Delta{|\varepsilon_0|} \right)^{1/4} |\varepsilon_0|
\end{equation}
for the width. The DoS at $E<0$ is the mirror image of Eq.\ (\ref{nuom3}). The total number of states in the two impurity bands equals $n_s$.
The DoS (\ref{nuom3}) coincides with the result derived in the clean limit [see Eq.\ (\ref{app:nuom3})] up to a factor of 4. This factor accounts for the difference in the definitions of DoS. [The DoS (\ref{app:nuom3}), defined for Bogoliubov quasiparticles existing only at $E>0$, corresponds to Eq.\ (\ref{nuom3}) (electron component) folded together with its mirror image (hole component) and, additionally, summed over the spin projections.]

As the concentration of magnetic impurities (and hence $\gamma_s$) grows, the upper edge of the band merges with the continuum at
$n_s/(\zeta\nu_0\Delta)=16\pi(2/\sqrt{3}-1)^{3/2}\approx 3.06$ [this result can be obtained from Eq.\ (\ref{u_eq}) and differs only numerically from the condition obtained for the case of polarized magnetic impurities in Sec.~\ref{subsec:imp-band}].
At higher concentrations the AG regime is realized, with the gap width reduced due to the scattering off magnetic impurities. As follows from Eq.\ (\ref{u_eq}), the quasiparticle continuum starts at energy lower than the single-impurity bound-state energy $E_0$, and the gap $E_g$ is reduced progressively with the increase of $\gamma_s$. At $\gamma_s^{2/3} \gg \zeta^2$, one may replace $u^2-\epsilon_0^2$ with $u^2-1$ in Eq.\ (\ref{u_eq}). Then it simplifies to the form
\begin{equation} \label{u_AG}
u \left( 1-\gamma_s \frac 1{\sqrt{1-u^2}} \right) = \frac E\Delta,
\end{equation}
exactly as considered by Abrikosov and Gor'kov.\cite{AG} Note that $\gamma_s$ itself can still be small, and we mainly focus on that case:
\begin{equation} \label{gammas}
\gamma_s\ll 1.
\end{equation}

The AG regime, corresponding to a single gap in the spectrum, has two characteristic features, essential for the low-frequency dissipation: first, the gap is suppressed (the relative scale is determined by $\gamma_s^{2/3}$), and second, the square-root BCS singularity of the DoS at $E=\Delta$ is smeared.
The gap $E_g$ in units of $\Delta$ is\cite{AG}
\begin{equation} \label{epsilong}
\frac{E_g}\Delta = \left( 1-\gamma_s^{2/3} \right)^{3/2} \approx 1-\frac 32 \gamma_s^{2/3}.
\end{equation}
Solving the Usadel equation at energies close to the actual gap, $(E -E_g) \ll \gamma_s^{2/3} \Delta$, we obtain\cite{Sri_comment}
\begin{equation} \label{neargap}
G(E) = -\frac i{\gamma_s^{1/3}} + \frac 1{\gamma_s^{2/3}} \sqrt{\frac{2(E-E_g)}{3\Delta}}.
\end{equation}
The square-root behavior of the DoS [see Eq.\ (\ref{DoS_u})] in the vicinity of the gap was found in Refs.~\onlinecite{Skalski,Maki}.

At low concentration, magnetic impurities substantially modify the spectrum in a relatively narrow domain of energies, $|E -\Delta|\lesssim \gamma_s^{2/3} \Delta$. We take $E=\Delta$ as a reference point within the domain, and find $G$ and its derivative with respect to energy:
\begin{gather}
G(\Delta) = \frac 1{(2i)^{1/3} \gamma_s^{1/3}}, \label{at1a} \\
G'(\Delta) = \frac i{3\gamma_s} + \frac{\sqrt 3 -i}{3\cdot 2^{4/3} \gamma_s^{1/3}}.
\end{gather}
The maximum of the DoS is achieved slightly above $E=\Delta$. At the same time, considering the derivative $G'(\Delta)$ and the BCS solution that makes it possible to approach the maximum from the side of higher energies, we conclude that the maximal DoS is of the same order as the value determined by Eq.\ (\ref{at1a}). This result was found in Ref.~\onlinecite{LO_1971}.

\begin{figure}[t]
 \includegraphics[width=\hsize]{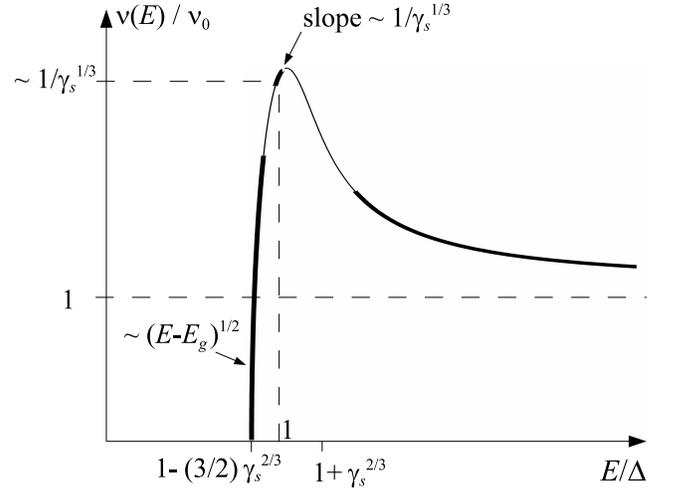}
\caption{Schematic (not numerically exact) DoS for a superconductor in the AG regime with a small concentration of magnetic impurities. The bold sectors of the curve are known analytically.}
 \label{fig:DoS}
\end{figure}

The results for the DoS are summarized in Fig.~\ref{fig:DoS}. Note that the square-root behavior of the DoS near the gap edge [following from Eq.\ (\ref{neargap})], being applied at $E=\Delta$ (which is already beyond its applicability range), yields $\nu(\Delta) = \gamma_s^{-1/3}$, which only slightly differs from the accurate result $\nu(\Delta) = (\sqrt 3/ 2^{4/3}) \gamma_s^{-1/3} \approx 0.7 \gamma_s^{-1/3}$ following from Eq.\ (\ref{at1a}).

In the evaluation of dissipation with the help of Eq.\ (\ref{dissip}), a good approximation for the DoS is given by
\begin{equation} \label{DoS}
\Re G(E) \simeq \left\{
\begin{array}{ll}
\frac 1{\gamma_s^{2/3}} \sqrt{\frac{2(E-E_g)}{3\Delta}}, & E_g \lesssim E \lesssim \Delta, \\
\frac{\sqrt 3}{2^{4/3} \gamma_s^{1/3}}, & \Delta \lesssim E \lesssim \Delta (1+\gamma_s^{2/3}), \\
\frac E{\sqrt{E^2-\Delta^2}}, & \Delta (1+\gamma_s^{2/3}) \lesssim E.
\end{array}
\right.
\end{equation}
At the matching points, the above expressions match by the order of magnitude, being different only by numerical factors close to 1. Therefore, this is a rather accurate approximation for calculating the dissipation up to order-of-one numerical factors (at the same time, in the limiting cases that we consider below, the numerical factors will be asymptotically exact).

At small $\zeta$ and $\gamma_s$, magnetic impurities significantly affect the electron spectrum only at energies close to $\Delta$. In that energy range [corresponding to Eq.\ (\ref{nuom3}) and to the first two lines in Eq.\ (\ref{DoS})], we have $u\approx 1$ and hence $F\approx iG$ [as follows from Eq.\ (\ref{tEtD})], so that finally $\Im F \approx \Re G$. Outside that domain, the functions $F$ and $G$ in the leading order have the conventional BCS form [with $\Im F(E) \approx \Delta/\sqrt{E^2-\Delta^2}$ instead of the last line in Eq.\ (\ref{DoS})].

Note that we have discussed $\Re G(E)$ and $\Im F(E)$ at $E>0$. At the same time, $\Re G$ is even while $\Im F$ is odd with respect to $E$.

\subsection{Dissipative conductivity} \label{sec:dissip}

Having found the Green functions, we are ready to calculate the dissipation. The expression for the response kernel of the current to the external electromagnetic field is derived in Appendix~\ref{sec:app:kernel}. From the result for the kernel, we extract the expression for the dissipative part of conductivity:
\begin{multline} \label{dissip}
\frac{\sigma_1(\omega)}{\sigma_0} =
\frac 1\omega \int\limits_0^\infty d E \left( \tanh\frac{E_+}{2T} - \tanh\frac{E_-}{2T} \right) \\ \times \left[ \Re G(E_+) \Re G(E_-) + \Im F(E_+) \Im F(E_-) \right],
\end{multline}
where $E_\pm=E\pm\omega/2$ and $\sigma_0 =2e^2 \nu_0 D$ is the Drude conductivity.
This formula is the dirty-limit counterpart of Eqs.\ (\ref{abspow2}) and (\ref{abspow3}) for the absorption power in the clean limit.
The combination of the Green functions $G$ and $F$ here replaces the product of the quasiparticle densities of states and the combinations of the coherence factors in Eqs.\ (\ref{abspow2}) and (\ref{abspow3}).

The analysis of the present section is be based on Eq.\ (\ref{dissip}).
At the same time, the results can be easily converted into the language of the surface impedance, as discussed in Sec.~\ref{sec:app:impedance}. Small but finite dissipative conductivity implies a small but finite real part of the surface impedance.

\subsubsection{$T=0$: Transitions across the gap} \label{sec:dissip_zeroT}

At zero temperature the dissipation is absent below a certain threshold frequency. In the absence of magnetic impurities, $\omega$ must exceed $2\Delta$ so that a Cooper pair can be broken and two quasiparticles are created above the gap. At frequencies slightly above the threshold, $(\omega-2\Delta) \ll \Delta$, the dissipative part of conductivity is\cite{MB,AGKh,Nam,Abrikosov}
\begin{equation}
\frac{\sigma_1(\omega)}{\sigma_0} = \frac\pi 2 \left( \frac\omega{2\Delta} -1 \right).
\end{equation}

In the presence of magnetic impurities, the BCS singularity of the Green functions is smeared out and hence the threshold behavior of the dissipation changes. The threshold frequency in the AG regime is $2E_g$, and the nonzero contribution to the integral in Eq.\ (\ref{dissip}) comes only from the region of energies $0<E<(\omega/2-E_g)$, where the arguments of the Green functions are close to $\pm E_g$ so that, at first sight, we can employ the square-root form for $\Re G$ and $\Im F$ [the first line in Eq.\ (\ref{DoS})] and take into account that $\Re G(E)$ is even while $\Im F(E)$ is odd. However, approximating $\Re G(E) = \Im F(E)$ at $E>0$ we find that the integrand in Eq.\ (\ref{dissip}) vanishes in the main order, so we should take into account a small difference between $\Re G$ and $\Im F$. This is easily done on the basis of the normalization condition $G^2 +F^2=1$. Note that it is the large absolute value of $G$ near $E_g$ [see Eq.\ (\ref{neargap})] that leads to $F\approx iG$ and hence to $\Im F \approx \Re G$. At the same time, the normalization condition immediately gives also a correction to this result:
\begin{equation} \label{ImFReG}
\Im F \approx \Re G - \Re \frac 1{2G}.
\end{equation}
Then we find nonzero dissipation which is quadratic in deviation from the threshold at $(\omega-2E_g)\ll \gamma_s^{2/3} \Delta$:
\begin{equation} \label{sgma}
\frac{\sigma_1(\omega)}{\sigma_0} = \frac\pi{6 \gamma_s^{2/3}} \left( \frac\omega{2E_g}-1 \right)^2.
\end{equation}
This result was obtain in Ref.~\onlinecite{LO_1971}. The quadratic threshold behavior of $\sigma_1$ is a consequence of the square-root threshold behavior of the DoS, and experimental results of Ref.~\onlinecite{DR} seem to confirm this $\sigma_1(\omega)$ dependence.

At lowest concentrations of magnetic impurities, the impurity band exists inside the gap, so that the threshold frequency for absorption is determined by transitions in which a Cooper pair is broken and two quasiparticles are created inside the impurity band. The threshold frequency for this process is $\omega_\mathrm{th} = 2(E_g-W/2)$, which is Eq.\ (\ref{omth}) written in different notation. Just above the threshold, at $(\omega-\omega_\mathrm{th})\ll W$, only the edge behavior (near an edge of the impurity band) of $\Re G$ and $\Im F$ enters Eq.\ (\ref{dissip}). According to Eq.\ (\ref{nuom3}), both functions near an edge grow as a square root. We again find a cancellation in the main order and have to employ Eq.\ (\ref{ImFReG}). The functional form is thus the same as in the AG regime, and the result differs from Eq.\ (\ref{sgma}) only by coefficients:
\begin{equation}
\frac{\sigma_1^{(b)}(\omega)}{\sigma_0} = \frac{2 \zeta^2 n_s^2}{\pi \nu_0^2 W^3 \omega_\mathrm{th}} \left( \omega -\omega_\mathrm{th} \right)^2.
\end{equation}

Next we consider the case of finite (but low) temperatures, when the threshold for absorption is absent since equilibrium thermally excited quasiparticles can absorb photons of arbitrarily small energy.

\subsubsection{Low $T$: Transitions within the AG states} \label{sec:AG}

We start with the consideration of the AG regime in the limit of small $\gamma_s$, so that quasiparticle states exist only above the gap $E_g$, which is weakly suppressed compared to $\Delta$, as described by Eqs.\ (\ref{gammas}) and (\ref{epsilong}). Most of the results of this subsection were obtained in previous publications, although sometimes in the framework of physically different problems and with the help of different techniques. Our consideration makes it possible to incorporate them in a unified manner.

In the limit that we consider, $\Re G(E)$ and $\Im F(E)$ entering Eq.\ (\ref{dissip}) are nonzero only above $E_g$ which is close to $\Delta$, so that at low frequencies and temperatures,
\begin{equation}
\omega,T \ll \Delta,
\end{equation}
two simplifications are possible: (a)~we can substitute the $\tanh$'s with their asymptotic exponential forms, and (b)~since the exponentials limit the integration to the region of $|E-\Delta|\lesssim T$, where $\Im F(E)$ is approximately equal to $\Re G(E)$, the contribution of the $(\Im F \Im F)$ term is the same as the contribution from the $(\Re G \Re G)$ term. Therefore, we can simplify Eq.\ (\ref{dissip}) as
\begin{multline} \label{sigmaGG}
\frac{\sigma_1(\omega)}{\sigma_0} = \frac 4\omega \int\limits_{E_g}^\infty dE \left( e^{-E/T} - e^{-(E+\omega)/T} \right) \\
\times \Re G(E) \Re G(E+\omega).
\end{multline}
At the same time, at $\omega\ll \Delta$ the external electromagnetic field cannot excite quasiparticles across the gap; therefore, the dissipation will be due to quasiparticles that are already excited due to low but finite temperature. It is thus natural to express the results for the dissipation in terms of the equilibrium quasiparticle density,
\begin{equation} \label{nqp}
n_\mathrm{qp} = 4 \nu_0 \int\limits_0^\infty dE \Re G(E) \frac 1{e^{E/T}+1},
\end{equation}
for which under the same assumptions ($T\ll \Delta$, $\gamma_s\ll 1$) we obtain two limiting cases:
\begin{align}
\frac{n_\mathrm{qp}}{\nu_0} &= \frac 2{\gamma_s^{2/3}} \sqrt{\frac{2\pi}{3\Delta}} T^{3/2} e^{-E_g /T},
&\text{at } T \ll \gamma_s^{2/3} \Delta,   \label{nqp2}\\
\frac{n_\mathrm{qp}}{\nu_0} &= 2 \sqrt{2\pi T \Delta} e^{-\Delta /T} ,
&\text{at } T \gg \gamma_s^{2/3} \Delta. \label{nqp1}
\end{align}

Below we consider various relations between the three energy scales $\omega$, $T$, and $\gamma_s^{2/3} \Delta$ (while all of them are much smaller than $\Delta$). Starting with the case of relatively strong magnetic scattering, $\omega,T \ll \gamma_s^{2/3} \Delta$, we note that, due to the exponentials, the main contribution to the integral in Eq.\ (\ref{sigmaGG}) comes from the narrow region above $E_g$, where the square-root asymptote for the DoS is valid [see the first line in Eq.\ (\ref{DoS})]. As a result,
\begin{equation} \label{K1}
\frac{\sigma_1(\omega)}{\sigma_0} = \frac{8T}{3\gamma_s^{4/3} \Delta} e^{-E_g/T} \sinh\left( \frac\omega{2T} \right) K_1 \left( \frac\omega{2T} \right),
\end{equation}
where $K_1$ is the Macdonald function (the modified Bessel function of the second kind).
Asymptotic forms of the Macdonald function at large and small values of the argument describe the two limiting cases for the $\omega/T$ ratio, so that finally at $\omega \ll T \ll \gamma_s^{2/3} \Delta$ we reproduce the result of Maki,\cite{Maki,Maki_comment}
\begin{equation} \label{dis6}
\frac{\sigma_1(\omega)}{\sigma_0}
= \sqrt{\frac 8{3\pi}} \frac 1{\gamma_s^{2/3}} \left( \frac{\Delta}T \right)^{1/2} \frac{n_\mathrm{qp}}{\nu_0 \Delta},
\end{equation}
while at $T \ll \omega \ll \gamma_s^{2/3} \Delta$ we find
\begin{equation} \label{dis3}
\frac{\sigma_1(\omega)}{\sigma_0}
= \sqrt{\frac 23} \frac 1{\gamma_s^{2/3}} \left( \frac\Delta\omega \right)^{1/2} \frac{n_\mathrm{qp}}{\nu_0 \Delta}.
\end{equation}

The result (\ref{K1}) was actually obtained by Larkin and Ovchinnikov in a physically different but formally equivalent problem.\cite{LO_1971}
They considered a superconductor with inhomogeneous effective interaction between electrons. For small-scale inhomogeneities the problem turns out to be mathematically equivalent to the case of homogeneous effective interaction in the presence of magnetic impurities. The results of Ref.~\onlinecite{LO_1971} thus apply to the problem considered in the present paper.

In the limit of weak magnetic scattering, $\omega, T\gg \gamma_s^{2/3} \Delta$, the main contribution to the integral in Eq.\ (\ref{sigmaGG}) comes from the region of energies above the smeared BCS singularity; in that region we can apply the unperturbed result for the DoS [see the last line in Eq.\ (\ref{DoS})]. We thus reproduce the results obtained in the absence of magnetic impurities\cite{AGKh,Nam,Abrikosov} (we have to take into account the correspondence noted by Nam\cite{Nam} between the extreme anomalous limit in the clean case\cite{AGKh} and the dirty case to which our calculations refer):
\begin{equation}
\frac{\sigma_1(\omega)}{\sigma_0} = \frac{4\Delta}{\omega} e^{-\Delta/T} \sinh\left( \frac\omega{2T} \right) K_0 \left( \frac\omega{2T} \right).
\end{equation}
Asymptotic forms of the Macdonald function $K_0$ at large and small values of the argument describe the two limiting cases for the $\omega/T$ ratio, so that finally at $\gamma_s^{2/3} \Delta \ll \omega \ll T$ we find
\begin{equation} \label{dis4}
\frac{\sigma_1(\omega)}{\sigma_0}
= \frac 1{\sqrt{2\pi}} \left( \frac\Delta T \right)^{3/2} \frac{n_\mathrm{qp}}{\nu_0 \Delta} \ln\frac{4T}{\gamma \omega}
\end{equation}
(where $\gamma \approx 1.78$ is Euler's constant), while at $\gamma_s^{2/3} \Delta \ll T \ll \omega$ the result is
\begin{equation} \label{dis1}
\frac{\sigma_1(\omega)}{\sigma_0}
= \frac 1{\sqrt 2} \left( \frac\Delta\omega \right)^{3/2} \frac{n_\mathrm{qp}}{\nu_0 \Delta}.
\end{equation}
The logarithmic divergence in Eq.\ (\ref{dis4}) at $\omega\to 0$ is due to the overlap of two square-root BCS singularities of the DoS.

Similarly to the case of Eq.\ (\ref{dis1}), at $T \ll \gamma_s^{2/3} \Delta \ll \omega$ the second exponential in Eq.\ (\ref{sigmaGG}) can be neglected while the $\omega$-shifted DoS is nearly constant, $\Re G(E+\omega) \approx \sqrt{\Delta/2\omega}$, in the essential region of integration. As a result, the remaining integral in both cases is exactly the same as the one determining the quasiparticle density (\ref{nqp}) at $T\ll E_g$. Therefore, we obtain exactly the same result (\ref{dis1}) in terms of $n_\mathrm{qp}$, although the expressions for $n_\mathrm{qp}$ itself are different at $\gamma_s^{2/3} \Delta \ll T \ll \omega$ and $T \ll \gamma_s^{2/3} \Delta \ll \omega$ [Eqs.\ (\ref{nqp1}) and (\ref{nqp2}), respectively]. This result actually follows from Ref.~\onlinecite{LO_1971}.

Finally, at $\omega \ll \gamma_s^{2/3} \Delta \ll T$, relatively large temperature allows rather wide region of energy integration in Eq.\ (\ref{sigmaGG}), while the shift of the two DoS can be neglected. Similarly to the case of Eq.\ (\ref{dis4}), a logarithmic singularity appears in the integral; however, now it is cut off not due to the finite shift $\omega$ [as it was the case in Eq.\ (\ref{dis4})] but due to smearing of the square-root BCS singularity of the DoS by magnetic impurities. As a result, with the logarithmic accuracy we find
\begin{equation} \label{dis5}
\frac{\sigma_1(\omega)}{\sigma_0} = \frac 1{\sqrt{2\pi}} \left( \frac\Delta T \right)^{3/2} \frac{n_\mathrm{qp}}{\nu_0 \Delta}\ln\frac{T}{\gamma_s^{2/3} \Delta}.
\end{equation}

The results for $\sigma_1(\omega)$ at $\omega\ll \Delta$ are valid as long as the quasiparticle population may be described by the Boltzmann distribution with an effective temperature. The quasiparticle chemical potential should not necessarily be $0$ and the temperature $T$ may deviate from equilibrium as long as $T\ll\Delta$.

\subsubsection{Low $T$: Transitions within the impurity band} \label{sec:impband}

At equilibrium, the results for the dissipation $\sigma_1$ calculated in Sec.~\ref{sec:AG} for the AG regime in the limit of low temperatures always contain an exponentially small factor $\exp(-E_g/T)$, since the absorption of the electromagnetic field is due to the quasiparticles thermally excited above the gap. At the same time, as we discussed above, at low concentrations of magnetic impurities a band of quasiparticles states appears below the continuum. This means that at temperatures much smaller than the distance between the band and the continuum, the dominant contribution to the dissipation will be due to quasiparticles residing inside the impurity band. Below we consider this case, which is the main focus of our interest.

Our aim now is to calculate the dissipation (\ref{dissip}) for the case when it is due to transitions inside the impurity band; hence, $\omega < W$. At the same time, we assume the band to be narrow, so that its width, which can be written as $W = 4 \sqrt{\zeta \gamma_s} \Delta$, is much smaller than the distance from the center of the band $E_0$ to the bottom of the continuum $\Delta$: Since $E_0 = (1-2\zeta^2) \Delta$, the condition can be written as
\begin{equation}
\gamma_s^{1/2} \ll \zeta^{3/2}.
\end{equation}
The contribution of the impurity band is dominant at $T\ll \zeta^2 \Delta$, when the continuum contribution is exponentially smaller due to the $\exp(-2\zeta^2 \Delta/T)$ factor.

Since we consider energies of the order of $E_0$, that is, of the order of $\Delta$, and $\Re G(E) =\Im F(E)$ inside the narrow band (as discussed in Sec.~\ref{sec:DoS}), we can make the same simplifications that lead us to Eq.\ (\ref{sigmaGG}). The only difference is that now the region of the energy integration is confined to within the impurity band. Taking into account the explicit form of the DoS inside the band [Eq.\ (\ref{nuom3})] and making the variable of integration dimensionless, we obtain
\begin{multline} \label{sigmaGGb}
\frac{\sigma_1^{(b)}(\omega)}{\sigma_0} = \frac{4 n_s^2}{\pi^2 \nu_0^2 \omega W} e^{-\frac{E_0}T} \sinh\frac\omega{2T} \\
\times \int\limits_{-\left( 1-\frac\omega W \right)}^{1-\frac\omega W} d\epsilon\; e^{-\frac W{2T}\epsilon} \sqrt{1-\left( \epsilon -\frac\omega W \right)^2} \sqrt{1-\left( \epsilon +\frac\omega W \right)^2}.
\end{multline}
The integral contains the product of the two semicircle DoS (\ref{nuom3}) shifted by the frequency $\omega$. The overlap exists and produces nonzero dissipation only at $\omega <W$.

Similarly to Sec.~\ref{sec:AG}, we express the final results in terms of the equilibrium quasiparticle density (within the band). Calculating Eq.\ (\ref{nqp}) for the impurity band, we find
\begin{equation}
n_\mathrm{qp}^{(b)} = \frac{4 n_s T}W e^{-E_0/T} I_1\left( \frac W{2T} \right),
\end{equation}
where $I_1$ is the modified Bessel function of the first kind. In the two limiting cases for the $W/T$ ratio, we obtain
\begin{align}
n_\mathrm{qp}^{(b)} &= \frac{4}{\sqrt\pi} n_s \left( \frac TW \right)^{3/2} e^{-\frac{E_0-W/2}T}, &\text{at } T \ll W, \label{nqp2b}\\
n_\mathrm{qp}^{(b)} &= n_s e^{-E_0/T} , &\text{at } T\gg W. \label{nqp1b}
\end{align}

Below we consider various relations between the three energy scales, $\omega$, $T$, and $W$ (while all of them are much smaller than $\Delta$). Starting with the case of relatively large band width, $\omega,T \ll W$, we note that, due to the exponentials, the main contribution to the integral in Eq.\ (\ref{sigmaGGb}) comes from the narrow region above the lower limit of integration where the simplified square-root result for the DoS is valid: $\nu(E) \propto \sqrt{E-(E_0-W/2)}$. This situation is very similar to the case of Eq.\ (\ref{K1}) (cf.\ the discussion above that formula). As a result, we obtain the same functional form [with a different prefactor and with $(E_0-W/2)$ instead of $E_g$ in the exponent]:
\begin{equation}
\frac{\sigma_1^{(b)}(\omega)}{\sigma_0} = \frac{16 n_s^2 T}{\pi^2 \nu_0^2 W^3} e^{-\frac{E_0-W/2}T} \sinh\left( \frac\omega{2T} \right) K_1 \left( \frac\omega{2T} \right).
\end{equation}
Asymptotic forms of the Macdonald function at large and small values of the argument describe the two limiting cases for the $\omega/T$ ratio, so that finally at $\omega \ll T \ll W$ we find
\begin{equation} \label{eq:sigmab-dirty1}
\frac{\sigma_1^{(b)}(\omega)}{\sigma_0}
= \frac{4}{\pi^{3/2}} \frac{n_s n_\mathrm{qp}^{(b)}}{\nu_0^2 W^{3/2} T^{1/2}},
\end{equation}
while at $T \ll \omega \ll W$ the result is
\begin{equation} \label{eq:sigmab-dirty3}
\frac{\sigma_1^{(b)}(\omega)}{\sigma_0}
= \frac{2}{\pi} \frac{n_s n_\mathrm{qp}^{(b)}}{\nu_0^2 W^{3/2} \omega^{1/2}}.
\end{equation}

At $(W-\omega)\ll T$ (while $W$ and $\omega$ are not necessarily close to each other), the exponential in Eq.\ (\ref{sigmaGGb}) can be replaced by unity in the region of integration, and the remaining integral can be written as
\begin{multline} \label{EK}
\frac{\sigma_1^{(b)}(\omega)}{\sigma_0} = \frac{16 n_s^2 (W+\omega)}{3\pi^2 \nu_0^2 W^2 \omega} e^{-\frac{E_0}T} \sinh\frac\omega{2T} \\
\times \left[ \left( 1+\frac{\omega^2}{W^2} \right) E(k) - \frac{2 \omega}{W} K(k) \right].
\end{multline}
Here $K$ and $E$ are the complete elliptic integrals of the first and second kind, respectively, and their argument is
\begin{equation}
k= \frac{W-\omega}{W+\omega}.
\end{equation}
In the limit of small frequency and high temperature, $\omega\ll W \ll T$, with the help of Eq.\ (\ref{nqp1b}), expression (\ref{EK}) can be simplified as
\begin{equation} \label{eq:sigmab-dirty2}
\frac{\sigma_1^{(b)}(\omega)}{\sigma_0} = \frac{8}{3\pi^2} \frac{n_s n_\mathrm{qp}^{(b)}}{\nu_0^2 W T}.
\end{equation}

There is an upper threshold for impurity band absorption at $\omega=W$. Close to that threshold, Eq.\ (\ref{EK}) gives $\sigma_1^{(b)}(\omega)\propto (W-\omega)^2 \theta(W-\omega)$, where $\theta$ is the Heaviside step function.

Comparing Eqs.\ (\ref{eq:sigmab-dirty1}), (\ref{eq:sigmab-dirty3}), and (\ref{eq:sigmab-dirty2}) in the dirty case with Eqs.\ (\ref{eq:sigmab-clean1}) and (\ref{eq:sigmab-clean2}) in the clean case, we notice that the dependences of the dissipative part of conductivity on $\omega$, $T$, and $W$ are the same. At the same time, the overall coefficients in the two cases are different due to different limits with respect to potential scattering.

\subsection{{Magnetic impurities as traps for nonequilibrium quasiparticles}}
\label{sec:traps}

At low temperatures, the density of equilibrium quasiparticles becomes negligible, and in reality $n_\mathrm{qp}$ is dominated by extraneously produced quasiparticles.\cite{aumentado} In quantum devices based on Josephson junctions, the adverse effect of quasiparticles on the coherence may be mitigated by the inclusion of traps --- sections of the superconducting leads with reduced value of the gap. We may pose a similar question with respect to the impedance: Can magnetic impurities reduce the dissipation caused by nonequilibrium quasiparticles? Such a possibility is most intriguing in the range of frequencies $\omega\lesssim |\varepsilon_0|$ [see Eq.\ (\ref{bs})] and quasiparticle densities $n_\mathrm{qp}\ll\zeta\nu_0\Delta$ (see Fig.~\ref{fig:Axis_ns}). In the following, we use the results of the two previous subsections to elucidate the dependence of the normalized dissipative conductivity $\sigma_1/\sigma_0$ on $n_s$.

For estimates, we assume the density of quasiparticles $n_\mathrm{qp}$ fixed, and their effective temperature low, $T\ll |\varepsilon_0| /\ln(\zeta\nu_0\Delta/n_\mathrm{qp})$, so that thermal ionization to the continuum does not prevent magnetic impurities from trapping the quasiparticles. In the absence of magnetic impurities, the dissipative response is defined by Eq.\ (\ref{dis1}). At small density, impurities reduce by $n_s$ the density of quasiparticles at the bottom of the continuous spectrum, yielding $\sigma_1/\sigma_0\approx (1/\sqrt{2})(\Delta/\omega)^{3/2}(n_\mathrm{qp}-n_s)/(\nu_0\Delta)$. In the interval of densities $n_\mathrm{qp}<n_s\ll\zeta\nu_0\Delta$, the impurity band may accommodate all the quasiparticles, while being too narrow to allow absorption of the field of frequency $\omega$. This low-absorption regime stretches up to the densities $n_s\lesssim (\pi/8)(\zeta\nu_0\Delta)(\omega/|\varepsilon_0|)^2$. At higher $n_s$, the impurity band is broad enough to allow for intraband absorption;
compared to the benchmark $n_s=0$ dissipation level, Eq.\ (\ref{eq:sigmab-dirty3}) yields $\sigma_1/\sigma_0$ smaller by a factor $\sim (\omega/|\varepsilon_0|) (n_s/\zeta\nu_0\Delta)^{1/4}$.
Upon further increase of $n_s$, the impurity band merges with the continuum. Deep in the AG regime, $n_s\gg\zeta\nu_0\Delta$, one may use Eq.\ (\ref{dis3}) for the estimate of dissipation; compared to the $n_s=0$ dissipation level, Eq.\ (\ref{dis3}) yields $\sigma_1/\sigma_0$ smaller by a factor $\sim (\omega/|\varepsilon_0|) (\zeta\nu_0\Delta/n_s)^{2/3}$.

To summarize the above estimates, in the selected region of frequencies, $\omega\lesssim |\varepsilon_0|$, and nonequilibrium quasiparticle densities, $n_\mathrm{qp} \ll\zeta\nu_0\Delta$, magnetic impurities indeed reduce the dissipative part of conductivity due to the reduction of the DoS available for the quasiparticle transitions.

\subsection{{Surface impedance}} \label{sec:app:impedance}

Finally, we reformulate the above results for the complex conductivity $\sigma(\omega)$ in terms of another important experimentally measurable quantity, the surface impedance:\cite{Abrikosov,Tinkham}
\begin{equation} \label{Z}
Z(\omega) = \sqrt{\frac{4\pi\omega}{i\sigma(\omega) c^2}}.
\end{equation}
The impedance can be written as $Z=R-iX$, where $R$ and $X$ are the surface resistance and reactance, respectively. The complex conductivity $\sigma = \sigma_1 +i\sigma_2$ is given by Eqs.\ (\ref{sigma}) and (\ref{QKeld}).

Since we always consider the case of weak dissipation, $\sigma_1 \ll \sigma_2$, we also obtain $R\ll X$. The response of the superconductor is then almost purely reactive with
\begin{equation}
X(\omega) =\sqrt{\frac{4\pi\omega}{\sigma_2(\omega) c^2}}.
\end{equation}
Under assumptions of Sec.~\ref{sec:dissip}, we can calculate $\sigma_2(\omega)$ in the simplest zero-temperature BCS model without magnetic scattering (since corrections due to low temperature and weak magnetic scattering, $T,\gamma_s^{2/3}\Delta \ll \Delta$, are negligible).
At the same time, the new qualitative feature arising due to finite $\sigma_1$ (calculated in Sec.~\ref{sec:dissip}), is the dissipation, i.e., small but nonzero $R$:
\begin{equation}
R(\omega) =\frac{X^3(\omega) c^2}{8\pi\omega} \sigma_1(\omega) .
\end{equation}

In the near-threshold situation of Sec.~\ref{sec:dissip_zeroT}, we are interested in frequencies $\omega\approx 2\Delta$, then\cite{Tinkham} $\sigma_2(2\Delta) = \sigma_0$, and we find
\begin{equation} \label{Rth}
R(\omega) = \frac{\sqrt{2\pi\Delta}}{c \sigma_0^{3/2}} \sigma_1(\omega).
\end{equation}

In the small-frequency limit considered in Secs.~\ref{sec:AG} and~\ref{sec:impband}, at $\omega \ll \Delta$, we find\cite{Tinkham} $\sigma_2(\omega) = (\pi\Delta/\omega) \sigma_0$ and finally obtain
\begin{equation} \label{Rlow}
R(\omega) = \frac{\omega^2}{\pi c (\sigma_0 \Delta)^{3/2}} \sigma_1(\omega).
\end{equation}

Thus, with the help of Eqs.\ (\ref{Rth}) and (\ref{Rlow}), the results for $\sigma_1(\omega)$ (the dissipative part of conductivity) calculated in Sec.~\ref{sec:dissip} are directly translated into the surface resistance.

\section{Conclusions}
\label{sec:concl}

We have employed the quasiclassical approach to study dissipation in superconductors due to small concentration of magnetic impurities. The superconductor was assumed to be in the dirty limit with respect to the potential (nonmagnetic) scattering. The concentration of magnetic impurities was assumed to be small enough so that the gap suppression is weak. Employing the extension (proposed by Marchetti and Simons\cite{MS}) of the Usadel equation beyond the Born approximation for magnetic scattering, we have considered in a unified manner both the AG regime (with a continuum of states above a gap) and the limit of lowest magnetic impurities' concentrations where the impurity band due to overlap of localized impurity states is formed below the edge of the continuum. The deviation from the Born limit was assumed to be finite but small, so that the subgap impurity states lie close to the continuum edge.

Our main results refer to the limit where the temperature $T$, the frequency $\omega$, and the gap suppression $\gamma_s^{2/3} \Delta$ are all much smaller than the BCS gap $\Delta$ in the absence of magnetic impurities. At the same time, the relation between $T$, $\omega$, and $\gamma_s^{2/3} \Delta$ was assumed to be arbitrary, and we have obtained explicit analytical expressions in various limiting cases.

Our results can be expressed in terms of quantities describing the response of superconductor to external electromagnetic field: the dissipative conductivity $\sigma_1$ (the real part of the complex conductivity $\sigma$) and the surface resistance $R$ (the real part of the surface impedance $Z$).

In the limit of small temperatures and small frequencies that we considered, at equilibrium, the dissipation is always proportional to the density of thermally excited quasiparticles and thus exponentially suppressed due to gapped character of the spectrum. On the other hand, fluctuations in positions of magnetic impurities can lead to finite DoS below the mean-field gap.\cite{LS,MS} Investigating the effect of the ``tail'' states on the dissipation is beyond the scope of our work. We only note that since the number of tail states is small (compared to the number of the ``mean-field'' states considered in this paper), there is a wide region of applicability for our results at subgap temperatures $T/\Delta \ll 1$, excluding only ultra-low temperatures.

\textit{Note added.} During preparation of this paper, we became aware of a preprint by Kharitonov \textit{et al.}\cite{Kharitonov}  addressing the same topic with a complementary approach. In that work the surface impedance of superconductors with magnetic impurities is evaluated numerically, with emphasis on the superconducting gapless regime having low energy states in the electron spectrum.

\begin{acknowledgments}
We are grateful to M.~H.\ Devoret, R.~J.\ Schoelkopf, M.~V.\ Feigel'man, and J.~S.\ Meyer for helpful discussions.
Ya.V.F.\ was supported by the RFBR (Grant No.\ 11-02-00077-a), the Russian Federal Agency of Education and Science (Contract
No.\ P799), and the program ``Quantum physics of condensed matter'' of the RAS.
This work was also supported by DOE Contract No.\ DE-FG02-08ER46482 (Yale) and
the Nanosciences Foundation at Grenoble, France. Ya.V.F.\ thanks Yale University for the hospitality. His visit was made possible by the gift of Victor and Marina Vekselberg to Yale University.
\end{acknowledgments}

\appendix

\section{Nonmagnetic disorder does not affect the electron spectrum}
\label{app:relation}

In this Appendix we demonstrate that the electron spectrum in the presence of magnetic impurities, regardless of their polarization, is insensitive to the nonmagnetic disorder. To see this we consider separately two different models. In Sec.~\ref{A:clean} we derive explicitly the equation determining the DoS in clean superconductors with randomly oriented spins and we establish the direct relation with the result obtained in Sec.~\ref{sec:dirty} (dirty superconductors, randomly oriented spins).
In Sec.~\ref{A:dirty} we formulate the Usadel equation (dirty limit) for superconductors with polarized magnetic impurities and find that the DoS corresponds to the one obtained in Sec.~\ref{sec:clean} (clean limit, polarized magnetic impurities).

\subsection{Magnetic impurities with randomly oriented spins} \label{A:clean}

In the clean limit, the Hamiltonian describing quasiparticles near the edge of the spectrum and interacting with randomly oriented weak  magnetic impurities is
\begin{multline}
\hat H = \sum_{\mathbf k, \sigma} \varepsilon_{\mathbf k} \alpha_{\mathbf k \sigma}^\dagger \alpha_{\mathbf k \sigma}
\\
+ J S\!\!\!\! \sum_{\mathbf k,\mathbf k',\sigma,\sigma',j}\!\!\!\!\alpha_{\mathbf k \sigma}^\dagger
(\hat{\mathbf s}_{\sigma\sigma'} \mathbf n_j) \alpha_{\mathbf k'\sigma'} e^{i(\mathbf k'-\mathbf k) \mathbf R_j},
\end{multline}
where $\hat{\mathbf s}$ is the vector of the Pauli matrices acting in the spin space and $\mathbf n_j$ is the (random) orientation of the spin of impurity $j$. [The Hamiltonian (\ref{Ham1}) for polarized magnetic impurities corresponds to all $\mathbf n_j=\hat z$.]
The $T$-matrix approximation for this model yields the self-consistency equation
\begin{equation} \label{grandom}
JS g = - \sqrt{\frac{|\varepsilon_0|}{-\varepsilon +\frac{n_s (JS)^2g}{1-(JSg)^2}}}
\end{equation}
for the momentum-integrated Green function $g(\varepsilon)$, which is now diagonal in spin space.
At low concentration of weak magnetic impurities ($\zeta\ll 1$ and $\gamma_s \ll1$) and for energies close to the edge of the continuous spectrum (i.e., $|u-1|\ll 1$), Eq.\ (\ref{grandom}) agrees with Eq.\ (\ref{u_eq}) that was derived in the dirty limit
provided one identifies $g=-2i\pi\nu_0\cos\theta$ [see also Eq.\ (\ref{tEtD})].
In particular, the analysis of Eq.\ (\ref{grandom}) in the limit $n_s\ll\zeta\nu_0\Delta$ yields a narrow impurity band with
\begin{equation} \label{app:nuom3}
\nu_B(E) = \frac{4n_s}{\pi W} \Re \sqrt{1 - \left( \frac{\varepsilon+|\varepsilon_0|}{W/2} \right)^2}
\end{equation}
for the total DoS (summed over quasiparticle spins) and
\begin{equation} \label{app:Wfin3}
W = 4\frac{1}{2^{1/4}\pi^{1/2}} \left(\frac{n_s}{\nu_0\Delta}\right)^{1/2} \left( \frac\Delta{|\varepsilon_0|} \right)^{1/4} |\varepsilon_0|
\end{equation}
for the width [which coincides with Eq.\ (\ref{Wfin3})].

\subsection{Spin-polarized magnetic impurities}
\label{A:dirty}

Following the derivation of Eq.\ (\ref{Usadel_mod}) by Marchetti and Simons,\cite{MS} we find that the Usadel equation for the case of polarized magnetic impurities is
\begin{equation} \label{Usadel_modup}
\frac D2 \nabla^2 \theta_\sigma + iE \sin\theta_\sigma + \Delta \cos\theta_\sigma
-\frac {i\sigma E_z\sin\theta_\sigma}{1-\zeta^2+2i\sigma \zeta\cos \theta_\sigma} =0.
\end{equation}
Here $\theta_\sigma$ parametrizes the quasiclassical Green functions that are different for electrons with spins up and spins down.
At low concentration of weak magnetic impurities
($\zeta\ll 1$ and $\gamma_s \ll1$) and for energies close to the edge of the continuous spectrum,
Eq.\ (\ref{Usadel_modup}) in a uniform superconducting state can be rewritten as
\begin{equation} \label{gsigma}
JS g_\sigma = - \sqrt{\frac{|\varepsilon_0|}{-\varepsilon +\sigma\frac{n_s JS}{1-\sigma JSg_\sigma}}}
\end{equation}
in terms of the spin-resolved momentum-integrated Green functions $g_\sigma=-2i\pi\nu_0\cos\theta_\sigma$.
The equation for down spin is the same as Eq.\ (\ref{g}) derived in the clean limit, while the one for spin up does not produce an impurity band.

\section{Derivation of the response kernel} \label{sec:app:kernel}

The Usadel equation in a superconductor reads
\begin{gather}
D\boldsymbol{\partial} (\check g\boldsymbol{\partial} \check g) - \hat\tau_3 \partial_t \check g - \partial_{t'} \check g \hat\tau_3 - [\hat\Delta, \check g] -i[\check\Sigma_s, \check g]=0, \label{eq:Usadel}\\
\hat\Delta = \begin{pmatrix} 0 & \Delta \\ \Delta^* & 0 \end{pmatrix},
\end{gather}
where the quasiclassical Green's function $\check g(\mathbf r,t,t')$ is a matrix in the Nambu (Pauli matrices $\hat\tau_i$) and Keldysh spaces (Pauli matrices $\hat\sigma_i$),
\begin{equation}
\check g = \begin{pmatrix} \hat g^R & \hat g^K \\ 0 & \hat g^A \end{pmatrix}_\mathrm{Keldysh},
\end{equation}
and the self-energy $\check\Sigma_s$ describes magnetic
scattering.\cite{Self} The covariant derivative $\boldsymbol{\partial}= \boldsymbol{\nabla}-i(e/c)[\hat\tau_3\mathbf A,.]$ depends on the vector potential $\mathbf A (\mathbf r,t)$ (we choose the gauge without the scalar potential). Time convolution is implicit in the matrix product.
The superconducting gap $\Delta(\mathbf r,t)$ enters the matrix (in the Nambu space) $\hat\Delta$ and solves the self-consistency equation
\begin{equation} \label{eq:self}
\Delta(\mathbf r,t) = -\frac{\pi |\lambda|\nu_0}8 \tr\left( \hat\sigma_2 \hat\tau_- \check g(\mathbf r,t,t) \right),
\end{equation}
where $\lambda$ is the BCS pairing constant.

We look for the solution to Eq.\ (\ref{eq:Usadel}) perturbatively in $\mathbf A$, expanding $\check g=\check g^{(0)}+\check g^{(1)}+\dots$ and $\Delta = \Delta^{(0)} + \Delta^{(1)} +\dots$ In the zeroth order, $\check g^{(0)}$ describes an equilibrium uniform superconductor.

In the next order,
\begin{equation}
\boldsymbol{\partial} \check g = \boldsymbol{\nabla} \check g^{(1)}- \frac{ie}c [\hat\tau_3\mathbf A, \check g^{(0)}],
\end{equation}
so that
\begin{equation}
\boldsymbol{\partial}(\check g\boldsymbol{\partial} \check g)
=
\check g^{(0)}\left(\boldsymbol{\nabla}^2 \check g^{(1)} -\frac{ie}c [\hat\tau_3 \boldsymbol{\nabla} \mathbf A, \check g^{(0)}]\right)
\end{equation}
in the same order. If we choose the London gauge with $\boldsymbol{\nabla} \mathbf A =0$ (accompanied by the requirement of $\mathbf{n A} =0$ at the surface), then the Usadel equation in the first order reads
\begin{equation}
\mathcal L \check g^{(1)} = [\hat\Delta^{(1)},\check g^{(0)}].
\end{equation}
The l.h.s.\ of this equation is a linear operator $\mathcal L$ acting on $\check g^{(1)}$, while the source term in the r.h.s.\ is proportional to $\Delta^{(1)}$. Taking the first order also in the self-consistency equation (\ref{eq:self}), we see that our choice of the gauge leads to $\Delta^{(1)}=0$ and $\check g^{(1)}=0$. (It is actually well known that $\Delta^{(1)} =0$ in the London gauge.\cite{AGD})

The current flowing in the superconductor reads
\begin{equation}
\mathbf j (\mathbf r,t) = \frac{\pi\sigma_0}{4e} \tr\left( \hat\sigma_1 \hat\tau_3 \check g\boldsymbol{\partial} \check g \right)_{\mathbf r,t,t}
\end{equation}
(where $\sigma_0 =2e^2 \nu_0 D$ is the Drude conductivity), which in the lowest order in $\mathbf A$ reduces to
\begin{equation}
\mathbf j (\mathbf r,t) = -\frac{i\pi\sigma_0}{4c} \tr\left( \hat\sigma_1 \hat\tau_3 \check g^{(0)} \hat\tau_3 \mathbf A \check g^{(0)} \right)_{\mathbf r,t,t}.
\end{equation}
Making the Fourier transform, we obtain
\begin{equation}
\mathbf j(\mathbf k,\omega) = - \frac 1c Q(\omega) \mathbf A(\mathbf k,\omega),
\end{equation}
with the response kernel [we omit the $(0)$ superscript of the Green function for brevity]
\begin{widetext}
\begin{equation} \label{Qom}
Q(\omega)
=\frac{i\sigma_0}{8}
\int dE \tr \left[
\hat\sigma_1 \hat\tau_3 \check g(E) \hat\tau_3 \check g(E-\omega)
\right]
=\frac{i\sigma_0}{8}
\int dE \tr \left[
\hat\tau_3 \hat g^R(E) \hat\tau_3 \hat g^K(E-\omega)+\hat\tau_3 \hat g^K(E) \hat\tau_3 \hat g^A(E-\omega)
\right].
\end{equation}
Note that independence of $Q$ on $\mathbf k$ implies locality of relation between $\mathbf j$ and $\mathbf A$ in the dirty limit.

The Keldysh component of the Green function can be expressed in terms of the retarded and advanced ones, while the advanced component can be written in terms of the retarded one,
\begin{equation}
\hat g^K (E) = \left( \hat g^R (E) -\hat g^A (E) \right) \tanh \frac E{2T},\qquad \hat g^A = -\hat\tau_3 \hat g^{R\dagger} \hat\tau_3,
\end{equation}
so that finally only the retarded component enters. In the case without the superconducting phase, it has the form
\begin{equation} \label{QR}
\hat g^R = G \hat\tau_3 + F \hat\tau_1
\end{equation}
[$G$ and $F$ are the functions introduced in Sec.~\ref{sec:dirty}, see Eq.\ (\ref{GFtheta})] and we obtain
\begin{align}
\frac{Q(\omega)}{\sigma_0} = \frac i2 \int\limits_{-\infty}^\infty dE \biggl\{ & \tanh\frac{E_-}{2T} \Bigl[
G(E_+) \Re G(E_-) - i F(E_+) \Im F(E_-) \Bigr] - \notag \\
- & \tanh\frac{E_+}{2T} \Bigl[ G^*(E_-) \Re G(E_+) + i F^*(E_-) \Im F(E_+)
\Bigr] \biggr\}, \label{QKeld}
\end{align}
\end{widetext}
where $E_\pm = E \pm \omega/2$
[note that the integrand in Eq.\ (\ref{QKeld}) is even with respect to $E$]. The difference of this expression for the kernel from the results of Refs.~\onlinecite{MB,AGKh,AGD,Nam,Skalski} is that our expression is written in terms of the \textit{quasiclassical} Green functions (in the dirty limit) that can be directly found from the Usadel equation. A similar expression was employed in Ref.~\onlinecite{LO_1971}, although it was formulated in terms of contour integrals in the plane of complex $E$ and the cuts in the plane were made from $\pm E_g$, the values of the AG gap; thus, the AG regime with a single gap was explicitly assumed. Our expression (\ref{QKeld}) is more general, applicable also in the case when the impurity band is present and in the case of gapless superconductivity.

Defining the complex conductivity\cite{Tinkham} $\sigma =\sigma_1+i\sigma_2$ that determines the (local) response of current to electric field,
\begin{equation}
\mathbf j(\mathbf k,\omega) = \sigma(\omega) \mathbf E(\mathbf k,\omega),
\end{equation}
we see that
\begin{equation} \label{sigma}
\sigma(\omega) = \frac{iQ(\omega)}\omega.
\end{equation}
Dissipation is determined by the real part of conductivity, $\sigma_1(\omega) = \Re \sigma(\omega) = -\Im Q(\omega)/\omega$, for which, according to Eq.\ (\ref{QKeld}), we obtain Eq.\ (\ref{dissip}) given in the beginning of Sec.~\ref{sec:dissip}.


\begin{thebibliography}{99}
\bibitem{twoimp}
The exponential in $R_{ij}$ factor of $|t_{ij}|$ may be obtained by the standard means relating $t_{ij}$ to the product of $\varphi (\mathbf r)$ and its derivative at the midpoint between the impurities. However, obtaining this way the preexponential factor is problematic because of the oscillatory nature of wave function (\ref{wf}). The short-range nature of the potential forming $\varphi (\mathbf r)$ allows us a full solution of the two-impurity problem. The main difference of the problem considered here from the corresponding one for a single quantum particle in the field of two short-range potentials\cite{Baz} stems from the structure of wave function (\ref{wf}) which stays finite at $r\to 0$.
\bibitem{footnote_sc}
The Usadel equation must be complemented by the self-consistency equation for the order parameter $\Delta$. As a result, $\Delta$ is suppressed by magnetic impurities. We do not treat the self-consistency equation explicitly, assuming $\Delta$ that we use to be the self-consistent order parameter corresponding to the given impurity parameters.
\bibitem{Sri_comment}
Alternatively, Eq.\ (\ref{u_AG}) can be rewritten as a fourth-order polynomial equation, and then an analytical solution is available, according to Ref.~\onlinecite{Nam} (see also the discussion of this solution in Ref.~\onlinecite{Sri}). However, this formal solution is bulky and nontransparent, therefore rather than simplifying it in the corresponding limiting cases, we choose to consider those cases directly.
\bibitem{Maki_comment}
In Sec.~VA2a of Ref.~\onlinecite{Maki}, Maki considers the ``low-temperature and low-frequency'' limit, which probably means that $T$ and $\omega$ are much smaller than both $\Delta$ and $\gamma_s^{2/3} \Delta$. The condition $\omega \ll T_c$ written above Eq.\ (102) in Ref.~\onlinecite{Maki} is probably a misprint for $\omega \ll T$. This would mean that Eq.\ (102) from Ref.~\onlinecite{Maki} is obtained at $\omega \ll T \ll \gamma_s^{2/3} \Delta < \Delta$, and in the limit when the last inequality becomes strong, we obtain agreement with our Eq.\ (\ref{dis6}).
\bibitem{Self}
The self energy for magnetic impurities with randomly oriented spins reads\cite{MS}
\begin{equation*}
\check\Sigma_s=
-\frac i{2\tau_s}
\frac 1{1+\zeta^2\hat\tau_3\check g\hat\tau_3\check g}
\hat\tau_3\check g\hat\tau_3.
\end{equation*}
For polarized magnetic impurities, the spin structure of the quasiclassical Green function $\check g$ needs to be taken into account (Pauli matrices $\hat s_i$) and the self energy reads
\begin{equation*}
\check\Sigma_s=E_z \frac 1{1+i\zeta\hat\tau_3 \hat s_3\check g} \hat\tau_3 \hat s_3.
\end{equation*}
\end{thebibliography}
\end{document}